\documentclass[proceedings]{rmaa}

\usepackage{rmaacite}

\usepackage{flushrt}

\renewcommand{\P}[1]{%
\ifnum#1=1\hbox{OW~168--326E}\fi
\ifnum#1=2\hbox{OW~167--317}\fi
\ifnum#1=3\hbox{OW~163--317}\fi
\ifnum#1=5\hbox{OW~158--323}\fi
\ifnum#1=0\hbox{OW~171--334}\fi}

\title{The Equilibrium Structure of Cosmological 
Halos: From Dwarf Galaxies to X-ray Clusters}
\author{Ilian T. Iliev 
and Paul R. Shapiro 
  \affil{University of Texas at Austin} }

\fulladdresses{ 
\item Ilian T. Iliev, 
Dept. of Physics, RLM 5.208, University of Texas, Austin, TX 78712
(iliev@astro.as.utexas.edu)
\item Paul R. Shapiro,
  Dept. of Astronomy, University of Texas, Austin, TX 78712
  (shapiro@astro.as.utexas.edu)
.}

\shortauthor{Iliev \& Shapiro}
\shorttitle{Equilibrium Structure of Cosmological 
Halos}


\abstract{A new model for the postcollapse equilibrium structure of virialized objects
 which condense out of the cosmological background universe is described 
and compared with observations and simulations of cosmological
halos. The model is based upon the assumption that virialized halos are isothermal, 
which leads to a prediction
of a unique nonsingular isothermal sphere for the equilibrium structure, 
with a core density which is proportional to the mean background density at 
the epoch of collapse. These predicted nonsingular isothermal spheres are in good
agreement with observations of the internal structure of dark-matter-dominated 
halos from dwarf galaxies to X-ray clusters. Our model also reproduces many of the 
average properties of halos in CDM simulations to good accuracy, suggesting that it is
a useful analytical approximation for halos which form from realistic initial conditions.
While N-body simulations find profiles with a central cusp, our nonsingular model
matches the simulated halos outside the innermost region well. This model may also
be of interest as a description of halos in nonstandard CDM models like self-interacting
dark matter, which have been
 proposed to eliminate the discrepancy between the cuspy halos
of standard CDM simulations and observed halos with uniform-density cores.}

\listofauthors{I.~T.~Iliev \& P.~R.~Shapiro}
\indexauthor{Iliev, I.~T.}
\indexauthor{Shapiro, P.~R.}

\begin{document}

\maketitle


\section{Introduction} 

The question of what equilibrium structure results when a density perturbation collapses 
out of the expanding background universe and virializes is central to the theory of
galaxy formation. The nonlinear outcome of the growth of Gaussian-random-noise 
cosmological density fluctuations due to gravitational instability in a hierarchical
clustering model like CDM is not amenable to direct analytical solution, however.
Instead, numerical simulations are required. As a guide to understanding these simulations, 
as a check on their accuracy, and as a means of extrapolating from simulation results 
of limited dynamic range, analytical approximations are nevertheless an
essential tool. One such tool of great utility has been the solution of the 
spherical top-hat perturbation problem (cf. Gunn \& Gott 1972, Padmanabhan 1993).
As used in the Press-Schechter (``PS'') approximation (Press \& Schechter 1974)
and its various refinements,
the top-hat model serves to predict well the number density of virialized halos of 
different mass which form at different epochs in N-body simulations.
An analytical model for the internal structure (e.g. mass profile, temperature, 
velocity dispersion, radius) of these virialized halos would be a further tool of great 
value for the semi-analytical modelling of galaxy and cluster formation, therefore.
Here we shall summarize our attempt along these lines.

Earlier work adopted crude approximations which used the virial theorem to match a 
collapsing top-hat perturbation either to a uniform sphere or to a singular isothermal sphere,
with the same total energy as the top-hat. Our first motivation, therefore, is simply 
to improve upon 
this earlier treatment by finding a more realistic outcome for the top-hat problem. As a 
starting point, we shall adopt the assumption that the final equilibrium is spherical,
isotropic, and isothermal, a reasonable first approximation to the N-body and gasdynamic
simulation results of the CDM model. As we shall see, the postcollapse analytical solution
derived from this assumption quantitatively reproduces many of the detailed properties
of the halos found in those simulations, so we are encouraged to believe that our
approximation is well justified. Our model is in disagreement, however, with the
N-body simulation result that, in their very centers, dark-matter-dominated halos have
cuspy profiles (e.g. Navarro, Frenk, \& White 1997; ``NFW''). 
By contrast, our model predicts a small, but uniform density core, as required
to explain the observed dwarf galaxy rotation curves and cluster mass profiles inferred 
from gravitational lensing. This discrepancy between the cuspy profiles of the N-body   
results and the observed dark-matter-dominated halos has led recently to a reexamination 
of the cold, collisionless nature of CDM, itself, and the suggestion that a variation
of the microphysical properties of the dark matter might make it more ``collisional'',
enabling it to relax dynamically inside these halos so as to eliminate the central cusp
(e.g. Spergel \& Steinhardt 1999). While the details of this suggestion are still 
uncertain, our model serves to predict its consequences, to the extent that we are able 
to ignore the details of the relaxation process inside the halo and approximate the
final equilibrium as isothermal. In what follows, we shall describe our model and 
compare its predictions both with CDM simulation results and with observations of 
dwarf galaxy rotation curves and galaxy clusters. 

\section{The truncated isothermal sphere model}
Our model, as described in Shapiro, Iliev, and Raga (1999) for an Einstein-de~Sitter
universe and generalized to a low-density universe, either matter-dominated or
a flat one with a positive cosmological constant, in Iliev \& Shapiro (2000), 
is as follows: An initial top-hat density perturbation collapses and virializes, 
which leads to a truncated nonsingular isothermal sphere in 
hydrostatic equilibrium (TIS), a solution of the Lane-Emden 
equation (appropriately modified in the $\Lambda\neq 0$ case). Although the mass 
and total energy of the top-hat are conserved thru collapse and virialization,
and the postcollapse temperature is set by the virial theorem 
(including the effect of
a finite boundary pressure), the solution is not uniquely determined by
these requirements alone. In order to find a unique solution, some additional 
information 
is required. We adopt the anzatz that the solution selected by nature will be 
the ``minimum-energy solution'' such that the boundary pressure is that for 
which the 
conserved top-hat energy is the minimum possible for an isothermal sphere 
of fixed mass within a finite truncation radius. As a check, we appeal to
the details of the exact, self-similar, spherical, cosmological infall 
solution of Bertschinger (1985). In this solution, an initial overdensity
causes a continuous sequence of spherical shells of cold matter, both 
pressure-free dark matter and baryonic fluid, centered on the overdensity,
to slow their expansion, turn-around and recollapse. The baryonic infall
is halted by a strong accretion shock while density caustics form in the
collisionless dark matter, instead, due to shell-crossing. The postcollapse 
virialized object we wish to model is then identified with that particular
shock- and caustic-bounded sphere in this infall solution
for which the mass and total energy are the same as 
those of our top-hat before collapse and the trajectory of its 
outermost mass shell was identical to that of the outer boundary
of our collapsing top-hat at all times until it encountered the shock.
This spherical region of post-shock 
gas and shell-crossing dark matter in the infall solution  
is very close to hydrostatic and isothermal and has virtually the same 
radius as that of the minimum-energy solution for the TIS. This confirms
our ``minimum-energy'' anzatz and explains the dynamical origin of the 
boundary pressure implied by that solution as that which results from 
thermalizing the kinetic energy of infall. 

With this ``minimum-energy'' anzatz, we find that a top-hat perturbation 
collapse leads to a unique, {\bf nonsingular} TIS, which yields a universal, 
self-similar density profile for the  postcollapse equilibrium of cosmic 
structure. Our solution  has a unique length scale and amplitude set by the top-hat 
mass and collapse epoch, with a density proportional to the background density
at that epoch. The density profiles for gas and dark matter are
assumed to be the same (no bias). The final virialized halo has a flat density core.

Case I: matter-dominated cases, both flat and low density (see Fig. 1).
The core size is $r_0= 0.034\,\,\times$ radius $r_t$, where $r_t$ is the size of the
halo (i.e. truncation radius). The central density is $\rho_0= 514 \,\,\times$ surface density 
$\rho_t$. The 1D velocity dispersion $\sigma_V$ of the dark matter and the gas 
temperature $T$ are then given by $\sigma_V^2=k_BT/(\mu m_p)=4\pi G\rho_0r_0^2$.
[Note: this $r_0\equiv r_{0\rm,King}/3$, where $r_{0\rm,King}$ is
 the core radius defined as the ``King radius''
by Binney \& Tremaine (1987, equ. [4-124b])]. 
Compared to the standard uniform sphere (SUS) and singular isothermal 
sphere (SIS) approximations (see Table 1), the temperature is 
$T=2.16\,\, T_{\rm SUS}=0.72\,\, T_{\rm SIS}$.
At intermediate radii, $\rho$ drops faster than $r^{-2}$.

Case II: flat, $\Lambda\neq0$ models. The profile varies with epoch of collapse, 
approaching the universal shape of case I above for early collapse.
For example, for $\Omega_0=1-\lambda_0=0.3$ and $z_{\rm coll}=(0;0.5;1)$, we obtain
$r_t/r_0=(30.04;29.68;29.54)$, $\rho_0/\rho_t=(529.9;520.8;517.2)$, 
and $T/T_{\rm SUS}=(2.188;2.170;2.163)$, respectively.

\begin{figure}
\begin{minipage}[c]{0.5\textwidth}
\includegraphics[height=2.5in,width=4in]{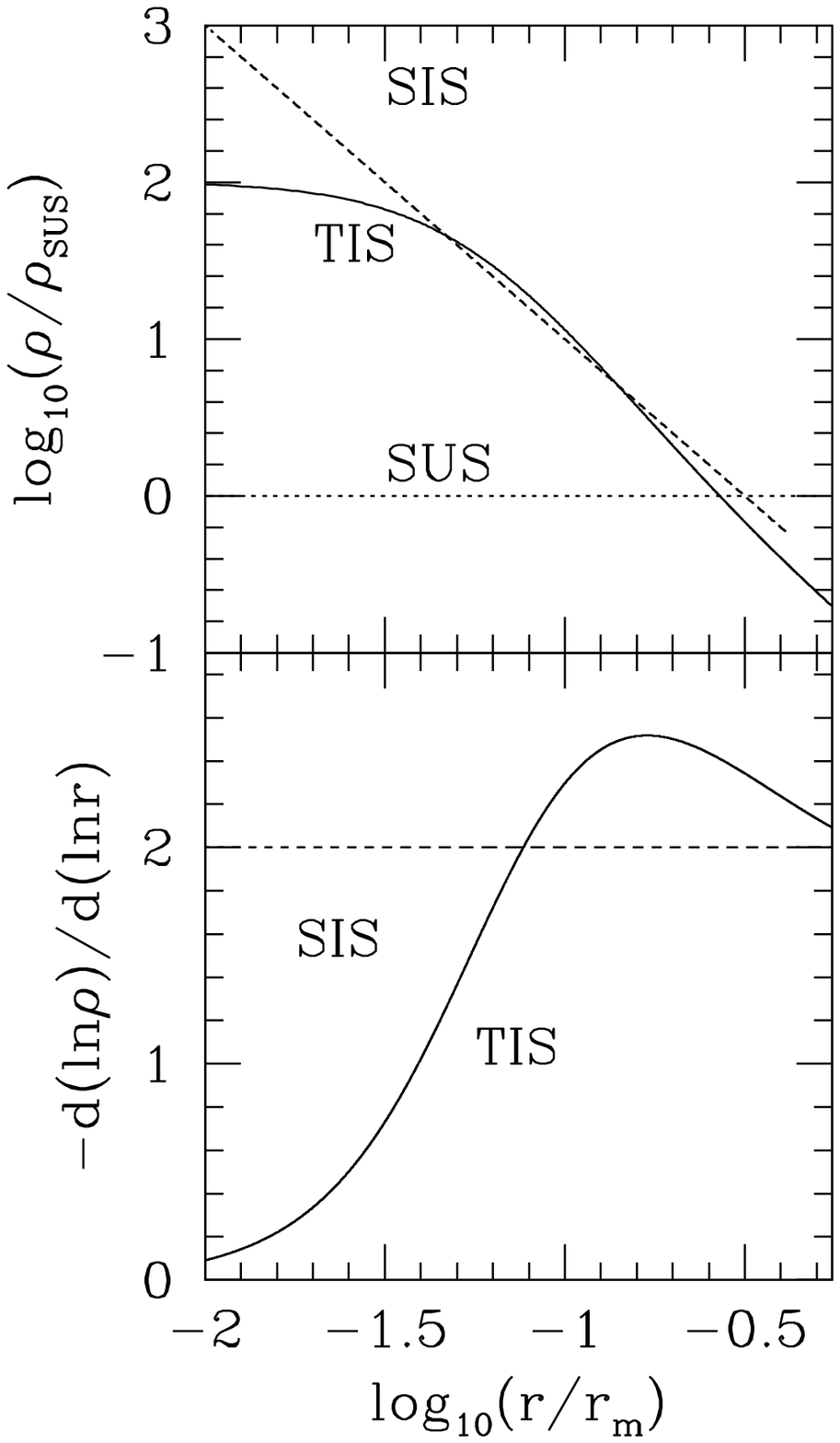}
    \label{profile}
\end{minipage}
\begin{minipage}[c]{0.2\textwidth}
\centering			
\begin{tabular}{@{}lcccc}
&&Table 1&\\ 
\hline
&SUS&SIS&TIS ($\Omega=1$)\\ \hline
$\eta=\frac{r_t}{r_m}$.....&0.5&0.417&0.554\\[2mm]
$\displaystyle{\frac{k_BT_{\rm vir}}
	{\left(\frac25\frac{GMm}{r_{\rm vir}}\right)}}$...&1&3&2.16\\
$\displaystyle{{\rho_0}/{\rho_t}}$...........& 1&$\infty$&514\\
$\displaystyle{{\langle\rho\rangle}/{\rho_t}}$...........&1&3&3.73\\
$\displaystyle{{r_t}/{r_0}}..............$& -- NA --&$\infty$&29.4\\
$\displaystyle{\frac{\langle\rho\rangle}{\rho_{crit}(t_{\rm coll})}}$.....&$18\pi^2$&
      $\displaystyle{18\pi^2\left(\frac65\right)^3}$
&130.5
\\
\hline
\end{tabular}
\end{minipage}
\vspace{-0.8cm}
\caption{Density profile of truncated isothermal sphere which forms from the 
virialization
of a top-hat density perturbation in a matter-dominated universe. Radius $r$ is in
units of $r_m$ - the top-hat radius at maximum expansion, while density $\rho$ is
in terms of the density $\rho_{SUS}$ of the standard uniform sphere approximation for 
the virialized, post-collapse top-hat. Bottom panel shows logarithmic slope of density 
profile. 
}
\end{figure}



\section{Dwarf Galaxy Rotation Curves and the 
$v_{max}-r_{max}$ correlations}

{\bf The TIS profile matches the observed mass profiles of 
dark-matter-dominated dwarf galaxies.} The observed rotation curves of
dwarf galaxies can be fit according to the following density profile
with a finite density core (Burkert 1995):
\begin{equation}
\rho(r)=\frac{\rho_{0,Burkert}}{(r/r_c+1)(r^2/r_c^2+1)}.
\end{equation}
The TIS profile gives a nearly perfect fit to the Burkert profile,
with best-fit parameters
$\displaystyle{{\rho_{0,Burkert}}/{\rho_{0,TIS}}=1.216}$,
${r_{c}}/{r_{0,TIS}}=3.134$ (see Fig.~2a).
\begin{figure}
	\centering
    \includegraphics[width=2.3in]{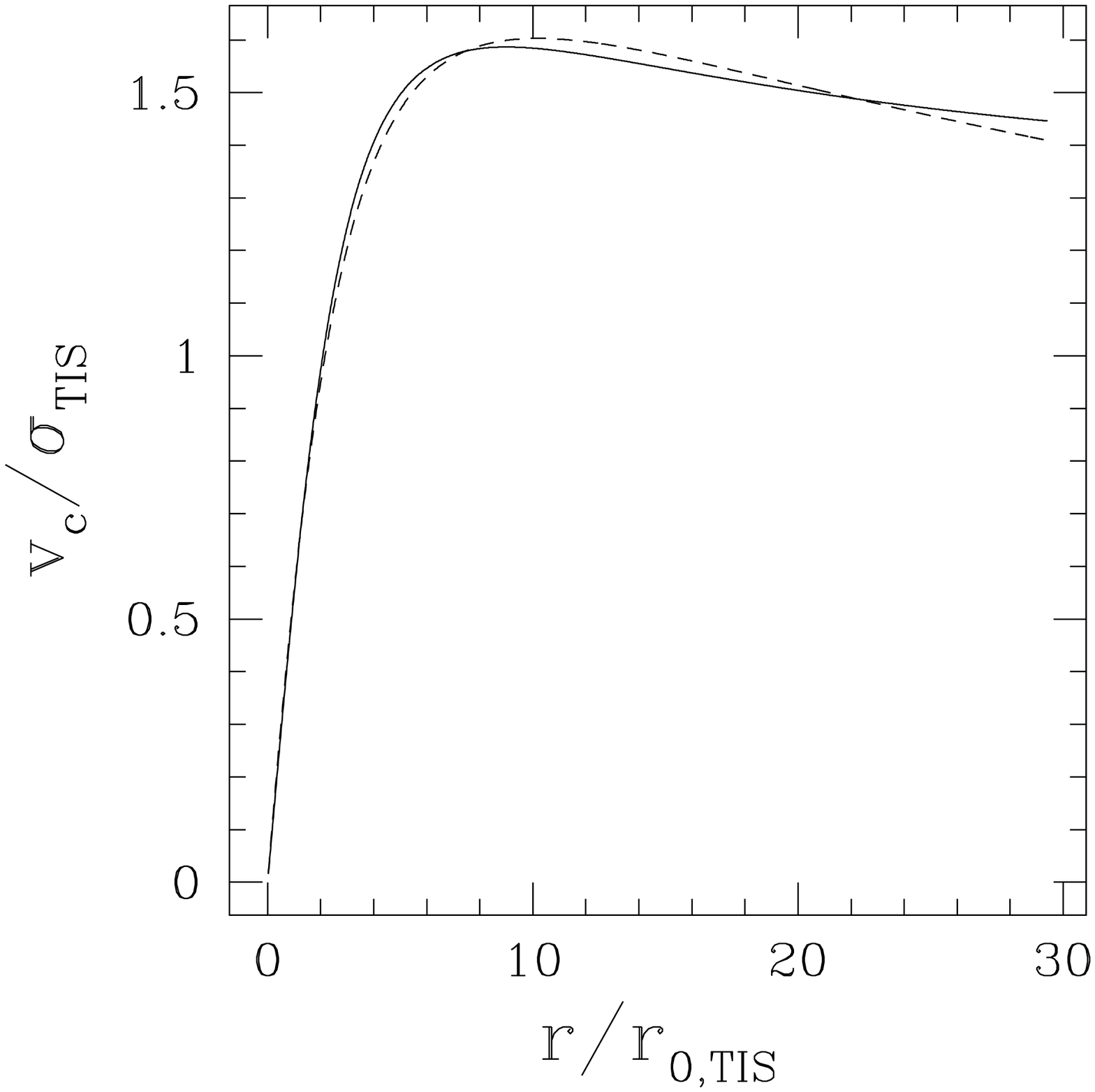}
	\hspace{1.5cm}
    \includegraphics[width=2.3in]{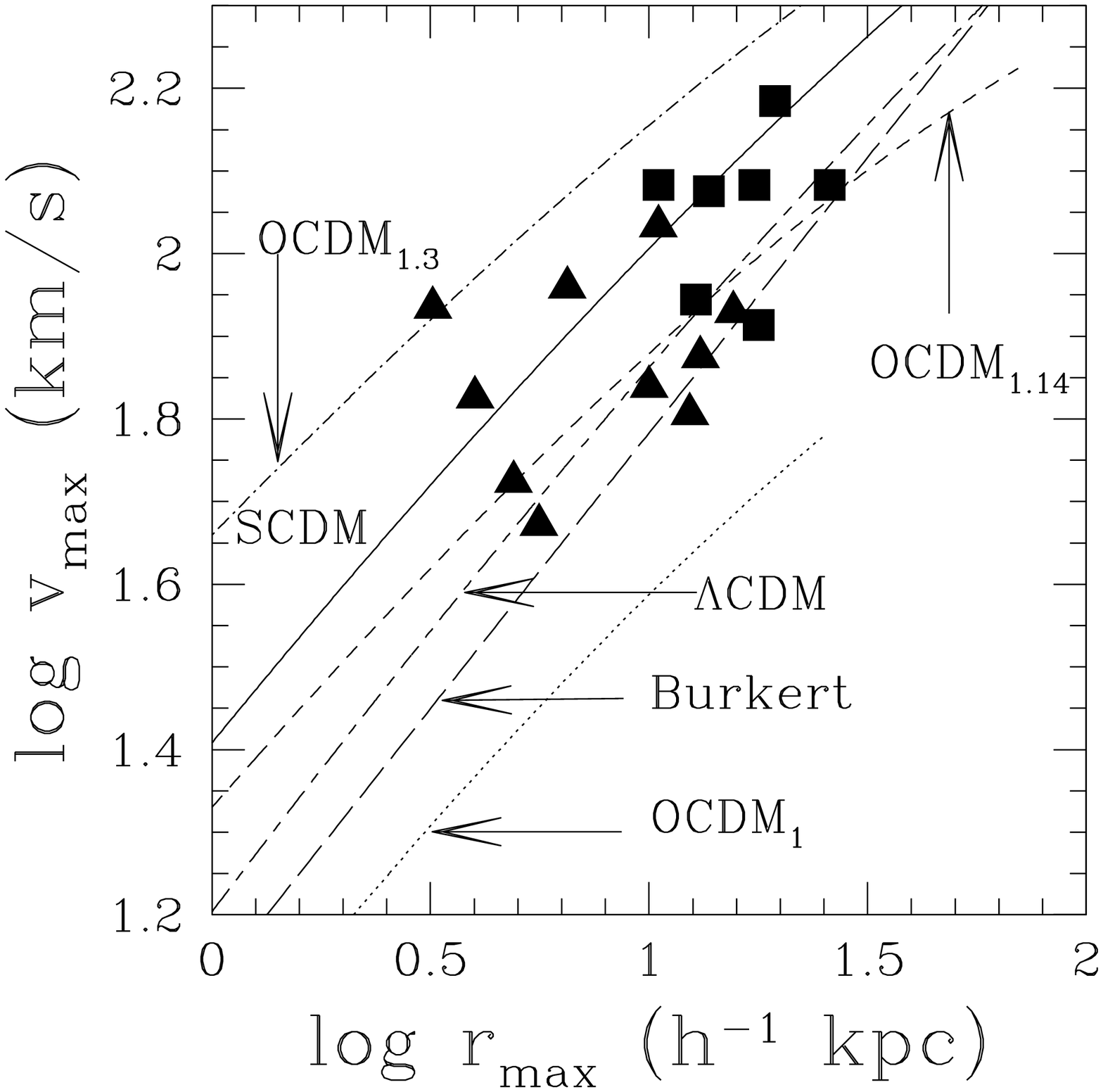}
    \caption{(a) (left) Rotation Curve Fit. 
Solid line = Best-fit TIS; Dashed line = Burkert profile, where
$\sigma_{TIS}^2=\langle v^2\rangle/3=k_BT/m$.
(b) (right) $v_{\rm max}$--$r_{\rm max}$ correlations.
 Observed dwarf galaxies (triangles) and LSB galaxies (squares) from Kravtsov et al. (1998);
Burkert: fit to data (Burkert 1995); SCDM: $\Omega_0=1$, $\lambda_0=0$, 
$\sigma_{8h^{-1}}=0.5$ (cluster-normalized); OCDM: $\Omega_0=0.3$, $\lambda_0=0$
(all COBE-normalized, subscript indicates $n$-value of tilt); 
$\Lambda$CDM: $\Omega_0=0.3$, $\lambda_0=0.7$, (COBE-normalized; no tilt);
$h=0.65$ for all.}
\end{figure}

How well does this best-fit TIS profile predict $v_{max}$, the maximum rotation
velocity, and the radius , $r_{\rm max}$, at which it occurs in the Burkert profile?
We find $\displaystyle{{r_{max,Burkert}}/{r_{max,TIS}}
	=1.13,\,}$ $\displaystyle{{v_{max,Burkert}}/{v_{max,TIS}}=1.01}$
(i.e. excellent agreement).

{\bf The TIS halo model explains the observed correlation of $v_{max}$ and $r_{max}$ 
for dwarf spiral and LSB galaxies},
when the TIS halo model is combined with the Press-Schechter model to 
predict the typical collapse epoch for objects of a given mass (i.e. the mass of 
the 1-$\sigma$ fluctuations vs. $z_{coll}$) (See Fig. 2b). Both of the flat, untilted 
CDM models plotted, cluster-normalized Einstein-de Sitter and COBE-normalized, flat,
low-density models ($\Omega_0=0.3$ and $\lambda_0=0.7$),
as well as the slightly tilted ($n=1.14$) open model ($\Omega_0=0.3$)
yield a reasonable agreement with the observed $v_{max}-r_{max}$ relation,
while the untilted ($n=1$) and strongly tilted ($n=1.3$) open models do not 
agree with the data. 

\section{Galaxy Halo $M-\sigma_v$ Relation}
{\bf Our TIS halo model predicts the
velocity dispersion of galactic halos of different mass which form in the CDM model
according to N-body simulations.} 
Antonuccio-Delogu, Becciani, \& Pagliaro (1999) used an N-body treecode 
at high-res ($256^3$ particles) to simulate galactic halos in regions of a 
single and of a double cluster. They found that the agreement with the TIS 
model is quite good, much better than with either
 of the other two models they considered, 
namely the singular isothermal sphere and the peak-patch model of Bond \& 
Myers (1996).
%

\section{Comparisons with Galaxy Cluster Observations and Simulations}
{\bf The TIS halo model predicts the internal structure of X-ray
clusters found by gas-dynamical/N-body simulations of cluster 
formation in the CDM model.} Our TIS model predictions agree astonishingly well
with the mass-temperature and the radius-temperature virial relations 
and integrated mass profiles derived from numerical simulations by Evrard, 
Metzler and Navarro (1996; ``EMN'').
Apparently, these simulation results are not sensitive to the discrepancy 
between our prediction of a finite density core and the N-body predictions 
of a density cusp for 
clusters in CDM. Let $X$ be the average overdensity inside the sphere of radius $r$,
$X\equiv{\langle\rho(r)\rangle}/{\rho_b}$. Then the
radius-temperature virial relation is defined as 
$r_X\equiv r_{10}(X)\displaystyle{\left( T/{10\, {\rm keV}}\right)^{1/2}}\,{\rm Mpc}$,
and the mass-temperature virial relation by
$M_X\equiv M_{10}(X)\displaystyle{\left( T/{10\, {\rm keV}}\right)^{1/2}}
	\,h^{-1} 10^{15}\,M_\odot$.
A comparison between our predictions of the mass - temperature relation
$r_{10}(X)$ and the results of EMN is given in Fig. 3a. 
\begin{figure}
	\centering
    \includegraphics[width=2in]{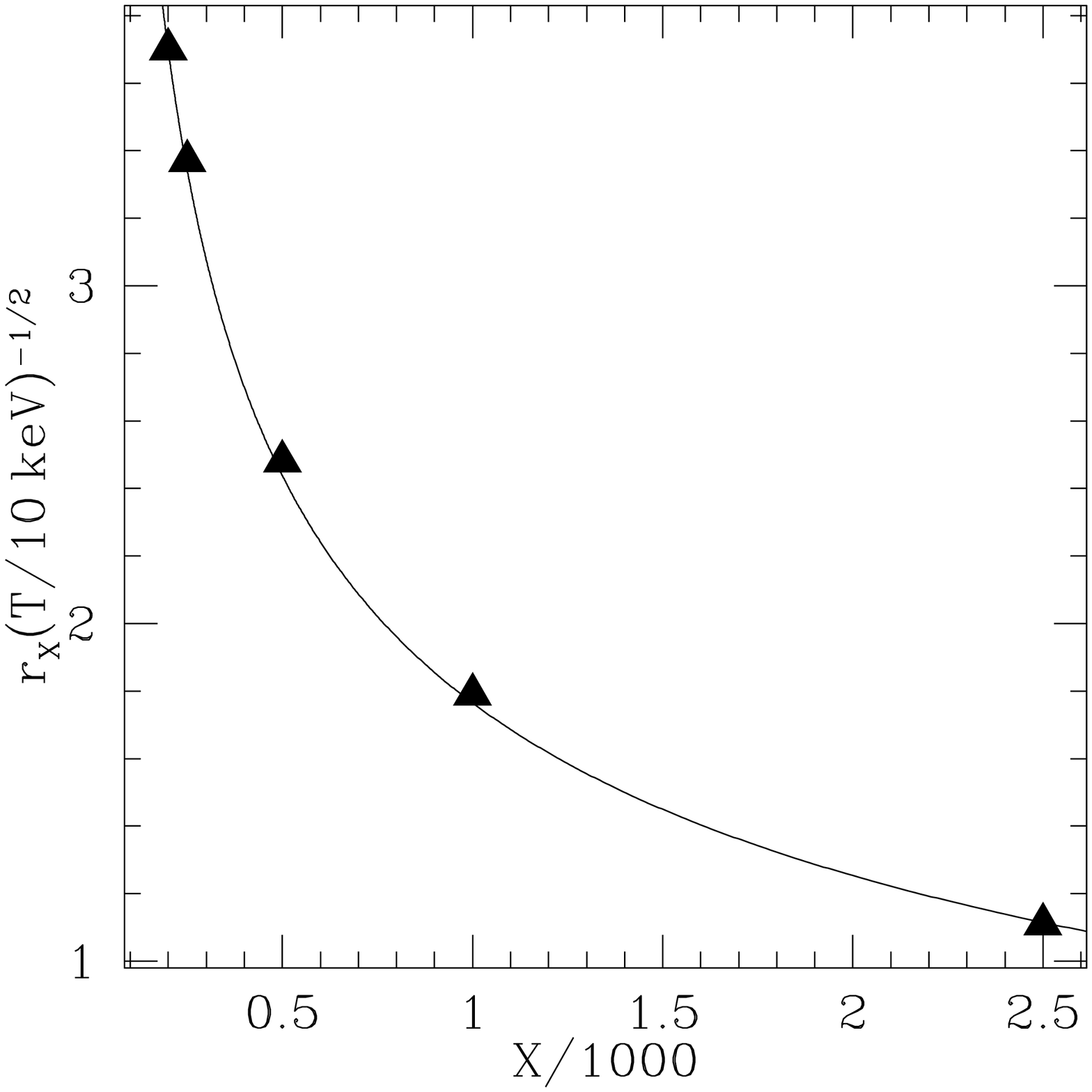}
	\hspace{1.5cm}
    \includegraphics[width=2in]{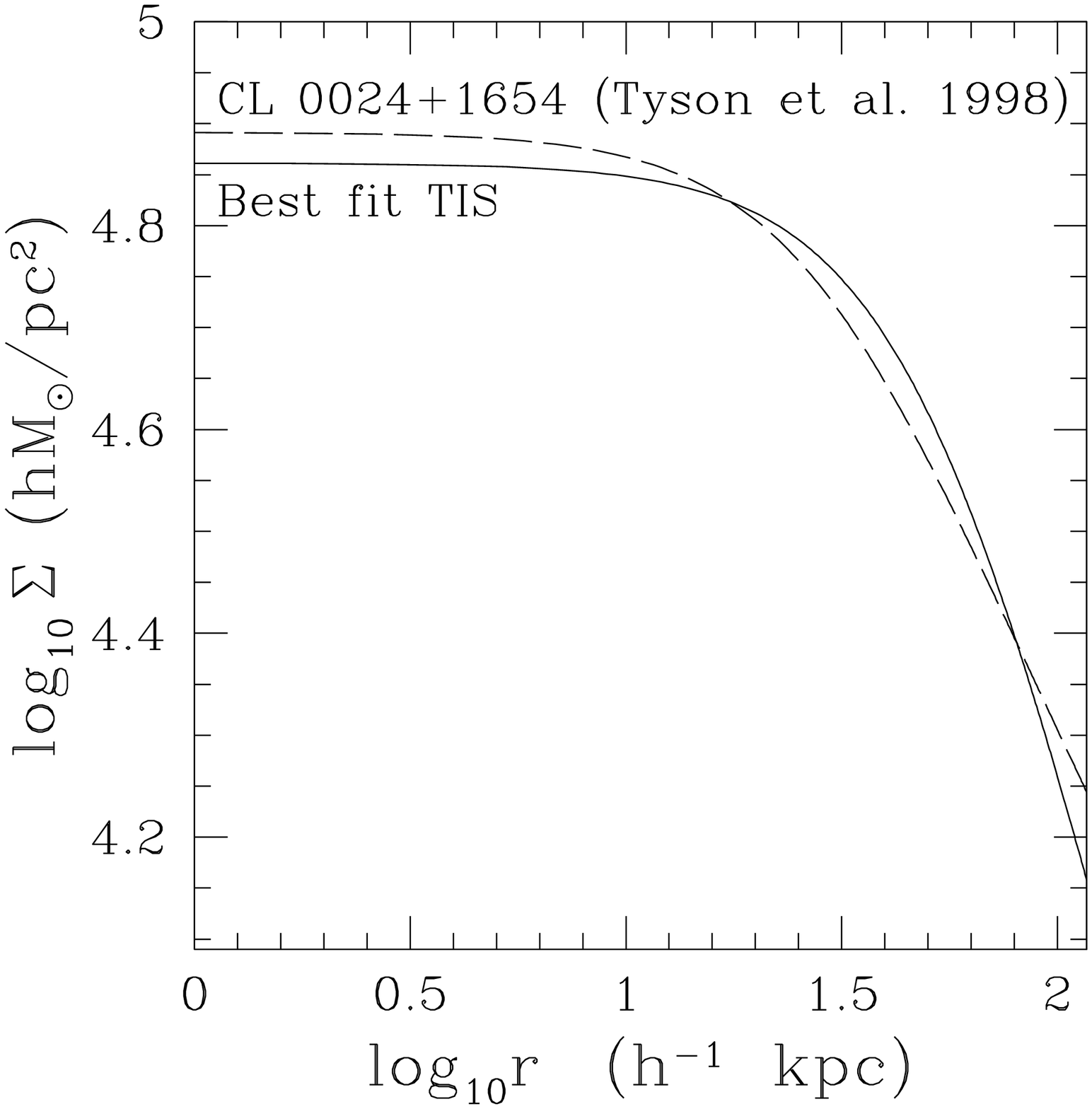}
    \caption{(a) (left) Cluster radius-temperature virial relation
($z=0$).  (triangles)  CDM simulation results as fit by 
	Evrard, et al. (1996);
(continuous line) TIS prediction.
(b) (right)
 Projected surface density of cluster CL 0024+1654 inferred from lensing 
measurements, together with the best-fit TIS model.}
    \label{burk_fit}
\end{figure}
For the mass-temperature virial relation
EMN obtain $M_{10}(500)=1.11\pm 0.16$ and $M_{10}(200)=1.45$, while our TIS solution
yields $M_{10}(500)=1.11$ and $M_{10}(200)=1.55$, respectively.

{\bf The TIS model for the internal structure of X-ray clusters predicts 
gas density profiles $\rho_{\rm gas}(r)$ and X-ray brightness profiles $I(\theta)$
which are well-fit by the standard $\beta-$profile,}
\begin{equation}
\rho_{gas}=\frac{\rho_0}{\displaystyle{\left(1+r^2/r_c^2\right)^{3\beta/2}}},\,\,\,
I=\frac{I_0}{\displaystyle{\left(1+\theta^2/\theta_c^2\right)^{3\beta-1/2}}},
\end{equation}
with $\beta$-values for the TIS $\beta$-fit which are quite close to 
those of simulated clusters in the CDM model but somewhat larger than the 
conventional observational result that $\beta\approx2/3$ (see tables below). 
However, recent X-ray results
suggest that the true $\beta$-values are larger than 2/3 when measurements at larger 
radii are used and when central cooling flows are excluded from the fit.

\hspace{-1cm}
\begin{minipage}[c]{0.15\textwidth}
\begin{tabular}{@{}lc}
{\bf $I(r)$ (observations)}&$\beta$\\\hline
Jones and Foreman (1999)& 0.4-0.8, ave. 0.6\\
Jones and Foreman (1992)& $\sim 2/3$\\
Balland and Blanchard (1997)& 0.57 (Perseus)\\
	&	0.75 (Coma)\\
Durret et al. (2000) & 0.53 (incl. cooling flow)\\
	&	0.82 (excl. cooling flow)\\ 
Vikhlinin, et al. (1999)&	0.7-0.8\\
(fit by Henry 2000)&\\
\hline
TIS $\beta$-fit ($r_c/r_{0,TIS}=2.639$)& 0.904\\\hline
\end{tabular}
\end{minipage}
\hspace{2.7in}
\begin{minipage}[c]{0.15\textwidth}
\centering
\begin{tabular}{@{}lc}
{\bf $\rho_{\rm gas}(r)$ (simulations)}&$\beta$\\\hline
Metzler and Evrard (1997) & 0.826 (DM)\\
&	 0.870 (gas)\\
Eke, Navarro, and Frenk (1998) & 0.82\\
Lewis et al. (1999) (adiabatic) &	$\sim 1$\\
Takizawa and Mineshige (1998) & $\sim 0.9$ \\
Navarro, Frenk, and White (1995)& 0.8\\
&\\[3mm]\hline
TIS $\beta$-fit ($r_c/r_{0,TIS}=2.416$) & 0.846\\
\hline
\end{tabular}
\end{minipage}

\hspace{4mm}
%
%
%
%
%
%

{\bf The TIS halo model can explain the mass profile with a flat density core 
measured by Tyson, Kochanski, and Dell'Antonio (1998) for cluster CL 0024+1654
at $z=0.39$, using the strong gravitational lensing of background galaxies by the 
cluster to infer the cluster mass distribution.}
The TIS model not only provides a good fit to the shape of the projected 
surface mass density distribution of this cluster within the arcs (see Fig. 3b), but
when we match the central value as well as the shape, our model predicts the overall 
mass, and a cluster velocity dispersion in close agreement with the value 
$\sigma_v=1200$ km/s measured by Dressler and Gunn (1992). By contrast, the NFW
fit which Broadhurst et al. (2000) reports can model the lensing data 
without a uniform-density core predicts $\sigma_V$ much larger than observed.

\section{Summary}
\begin{itemize}
\item The TIS profile fits dwarf galaxy rotation curves; combined with the
Press-Schechter approximation, it predicts the observed $v_{max}-r_{max}$ relation for dwarf 
and LSB galaxies.
\item The TIS predicted $M-\sigma_V$ relation 
agrees with high resolution N-body simulations of galactic halo formation by
Antonuccio-Delogu et al. (1999).
\item The predicted mass-radius-temperature scaling relations and integrated mass 
profile of the TIS model match simulation results 
for clusters in the CDM 
model in detail.
Our solution {\bf derives} the empirical fitting formulae of 
Evrard, et al. (1996), which also agree well with X-ray cluster 
observations at $z=0$.
\item The TIS X-ray brightness profile matches the $\beta$-fit profile with 
$\beta\approx0.9$, larger than typically reported by X-ray observers, but very
close to the results of
gas-dynamical/N-body simulations of X-ray clusters in the CDM model.
\item The TIS solution fits the cluster mass profile with uniform-density 
core derived from 
strong gravitational lensing data by Tyson et al. (1998) for CL 0024+1654 within
the arcs, while accurately predicting the observed $\sigma_V$ on larger 
scales, too. 
\end{itemize}


This work was supported by grants NASA NAG5-2785, NAG5-7363, and
NAG5-7821, NSF ASC-9504046, and Texas ARP 
3658-0624-1999, and a 1997 CONACyT National Chair of Excellence at UNAM for 
PRS.


\end{document}